\documentclass[aps,groupedaddress,showpacs,twocolumn,floatfix]{revtex4}
\usepackage{amsmath}
\usepackage{amssymb}
\usepackage{graphicx}
\usepackage{epsfig}

\begin{document}

\title{Frozen local hole approximation}
\author{Elke Pahl} 
\affiliation{Max-Planck-Institute for the Physics
of Complex Systems, N{\"o}thnitzer Stra{\ss}e 38, 01187 Dresden, Germany}
\author{Uwe Birkenheuer} 
\affiliation{Max-Planck-Institute for the Physics
of Complex Systems, N{\"o}thnitzer Stra{\ss}e 38, 01187 Dresden, Germany}

\date{\today}

\begin{abstract}
The frozen local hole approximation (FLHA) is an adiabatic approximation which
is aimed to simplify the correlation calculations of valence and conduction
bands of solids and polymers or, more generally, of the ionization potentials
and electron affinities of any large systems. Within this approximation
correlated {\it local} hole states (CLHSs) are explicitely generated by
correlating local Hartree-Fock (HF) hole states, i.~e.~($N$-1)-particle
determinants in which the electron has been removed from a {\it local}
occupied orbital. The hole orbital and its occupancy is kept frozen during
these correlation calculations, implying a rather stringent configuration
selection. Effective Hamilton matrix elements are then evaluated with the
above CLHSs; diagonalization finally yields the desired correlation
corrections for the cationic hole states. We compare and analyze the results
of the FLHA with the results of a full MRCI(SD) (multi-reference configuration
interaction with single and double excitations) calculation for two prototype
model systems, (H$_2$)$_n$ ladders and H--(Be)$_n$--H chains. Excellent
numerical agreement between the two approaches is found. Comparing the FLHA
with a full correlation treatment in the framework of quasi-degenerate
variational perturbation theory reveals that the leading contributions in the
two approaches are identical. In the same way it could be shown that replacing
a correlation calculation around a frozen {\it local} hole is in fact
equivalent, up to first order, to perform a much less demanding SCF
(self-consistent field) calculation around the hole to relax all other
occupied orbitals. Thus, both, the FLHA and the above SCF approximation, are
well-justified and provide a very promising and efficient alternative to fully
correlated wavefunction-based treatments of the valence and conduction bands
in extended systems.

\end{abstract}

\maketitle

\section{Introduction}

In the last decade, increasing interest in {\it wavefunction}-based
correlation methods for excited states in extended systems can be observed
\cite{GSF93,GSF97,GSB99,AFS00,AI01,A02,HVOB02,HPGB03,BB04,BBAF05}. This is
because one can rely on the very sophisticated numerical methods for molecules
which are well-established in quantum chemistry. These methods are
conceptually very clear and allow for systematic improvement: based on
self-consistent field calculations, electron correlation can be included
successively by considering determinants of increasing excitation order in
configuration interaction (CI) or related procedures. They yield approximate
correlated wavefunctions from which all quantities of interest can be derived.

Of course, wavefunction-based methods are computationally very demanding. In
order to arrive at manageable schemes for the description of solids or other
extended systems, local orbitals have to be introduced and the local character
of electron correlation must rigorously be exploited. Among these local
approaches \cite{P02} one can distinguish local MP2 (M{\o}ller-Plesset
perturbation theory up to 2$^{\rm nd}$ order) and coupled electron pair or
coupled cluster schemes \cite{SP87,FS95,HW96,SHW99,SB96,AKS01,PBCC05}, methods
which are based on a Green's function formalism \cite{A02,BBAF05} and
techniques which use local Hamiltonian matrix elements
\cite{GSF93,GSF97,AFS00,BB04} or other partitioning concepts
\cite{S92-1,S92-2,S92-3,PFS95,F02,WB04,ACCE03,KIANU99,FK04,MKAS05,BME05}.
Whereas the first-mentioned methods work in infinite systems and aim to
truncate the excitation space by an appropriate configuration selection
scheme, the effective Hamiltonian approach as well as the other partitioning
schemes assemble the correlation effects from finite subsystems (embedded
fragments) which are finally transfered to the infinite system. Most of these
methods focus on ground state properties, only the Green's function and local
Hamiltonian approaches are explicitely designed for excited states. In
particular, the local Hamiltonian approach has been used in the past quite
successfully to describe the correlation effects on the band structure of
covalent solids \cite{GSF93,GSF97,AFS00} and polymers \cite{BB04}. Yet, this
approach is quite involved and a more simpler, approximative way of computing
the local Hamiltonian has been proposed recently by our lab \cite{BFSpp}.

In this work we want to analyze the guiding approximations introduced in that
recent approach, and therefore address the question to which extent it is
possible to simplify the correlation calculations by focusing on ``frozen''
local hole configurations although, in reality, the electron hole is usually
delocalized over the entire system. Intuitively, this frozen local hole
approximation, is guided by the idea that the shape of the correlation
hole which is carried along by a traveling electron is essentially invariant
and is thus following the electron hole adiabatically. 

A related approach is the frozen core hole approximation \cite{RS03} for the
calculation of core-hole states where the core hole orbital is frozen and all
other orbitals are allowed to relax during an Hartree-Fock calculation. Such a
local view on core holes in bulk materials is also adopted in other
wavefunction-based investigations on excited core hole states
\cite{GBNB97,VHBB02}. While it seems quite natural to describe core holes in
that way, although in solids even core orbitals are Bloch waves, it is much
less obvious that a frozen local hole treatment should also be possible for
delocalized valence hole states.

In the following, we will outline the theoretical background of the frozen
local hole approximation (for valence holes) and then present and analyze the
numerical results obtained for two model systems, namely (H$_2$)$_n$ ladders
and H--(Be)$_n$--H chains. These two examples have been chosen to represent
the two cases of prevailing van der Waals and predominantly covalent
binding. Finally, an analytical, perturbative analysis of the frozen local
hole approximation is given.

\section{Theoretical background}

Starting point for our correlation calculations is the Hartree-Fock (HF)
ground state Slater determinant $|\Phi_0^N \rangle $ and energy $E_0^N$ of the
$N$-particle system. The canonical HF orbitals $|\nu \rangle $ with orbital
energies $\epsilon _{\nu} $ are divided into the $N/2$ energetically
lowest-lying orbitals which are occupied in $|\Phi_0^N \rangle $ and the
unoccupied, virtual orbitals. Furthermore, the occupied orbitals are
subdivided into core and valence orbitals depending on whether they are kept
frozen or are active during the subsequent correlation calculation. Within
these orbital groups we will switch between delocalized canonical orbitals and
localized orbitals, $\{ |\nu \rangle \}$ and $\{ |a\rangle \}$, using unitary
transformations, such as
\begin{equation}
|a\rangle = \sum_\nu | \nu \rangle U_{\nu a}
\end{equation}
for the occupied orbitals. In order to find the correlated cationic
($N$-1)-particle states of the system in mind, one first constructs the
so-called reference states $|\Phi_i^{N-1} \rangle$ of the system by removing
one electron out of one of the valence orbitals $|i\rangle $: $|\Phi_i^{N-1}
\rangle = \hat{c}_i |\Phi_0^N \rangle$. The so generated {\em canonical hole}
states $|\Phi_{\nu}^{N-1} \rangle$ and {\em local hole} states $|\Phi_a^{N-1}
\rangle$ are connected via
\begin{equation}
  \label{trafo} |\Phi_{\nu}^{N-1} \rangle = \sum_a |\Phi_a^{N-1} \rangle
  (U^{-1})_{a\nu }^\ast =\sum_a |\Phi_a^{N-1} \rangle U_{\nu a} \quad
\end{equation} 
Since we are only dealing with cationic states here we will omit the upper
index ($N$-1) in the following. The {\it canonical} hole states are the eigenstates
of the CI matrix in the reference space. They describe the hole states of the
system according to Koopmans' theorem \cite{K33}.

Including correlation effects one arrives at correlated wave functions
$|\Psi_{\nu} \rangle $ with energies $E_{\nu}$. In analogy to
Eq.~(\ref{trafo}) we then define so-called {\em correlated local hole states}
$|\Psi_a \rangle $ (see Fig.~\ref{figholestates})
\begin{equation} 
  \label{corrlocstates}
  |\Psi_a\rangle := \sum_{\nu} |\Psi_{\nu}\rangle U_{\nu a}^\ast 
\end{equation} 
and an effective Hamilton matrix $\underline{\underline{H}}^{\rm eff} =
(H_{ab}^{\rm eff})$ with 
\begin{equation} 
  \label{effham} H_{ab}^{\rm eff} := 
  \langle \Psi_a | \hat{H} | \Psi_b \rangle = 
  \sum_{\nu} U_{\nu a} E_{\nu} U_{\nu b}^\ast
  \quad.
\end{equation} 
The effective Hamiltonian $\hat{H}^{\rm eff}$ is constructed such that
diagonalizing its local matrix representation $\underline{\underline{H}}^{\rm
eff}$ precisely yields the desired correlated energies $E_{\nu}$ of the
cationic states. Thus, provided one knows the correlated local hole states
$|\Psi_a\rangle$ defined in Eq.~(\ref{corrlocstates}), one can find the
cationic energies by simply solving a small eigenvalue problem.

\setlength{\unitlength}{1cm}
\begin{figure}
\begin{picture}(6,3.5)
\put(0.5,3){$\{ |\Phi_{\nu} \rangle \} \quad \stackrel{U}{\longleftrightarrow}
\quad \{ | \Phi_{a} \rangle \} \quad \quad $ HF states}
\put(0.5,1){$\{ |\Psi_{\nu} \rangle \} \quad \stackrel{U}{\longleftrightarrow}
\quad \{ | \Psi_{a} \rangle \} \quad \quad $ correlated states}
\put(0.9,2.5){\vector(0,-1){0.8}}
\put(1.1,2.1){{\scriptsize MRCI}}
\multiput(3.3,2.1)(0,0.2){3}{\line(0,-1){0.15}}
\put(3.3,1.9){\vector(0,-1){0.15}}
\put(3.5,2.1){{\scriptsize FLHA}}
\end{picture}
\caption{Relation between the delocalized hole state on the Hartree-Fock
level $|\Phi_{\nu}\rangle$ and the corresponding correlated states
$|\Psi_{\nu}\rangle$, and their localized counterparts, $|\Phi_{a}\rangle$ and
$|\Psi_{a}\rangle$, respectively. Within the frozen local hole approximation
(FLHA) approximate correlated local hole states $|\tilde{\Psi}_a\rangle$ are
constructed without explicit reference to the true correlated states
$|\Psi_{\nu}\rangle$.}
\label{figholestates}
\end{figure}
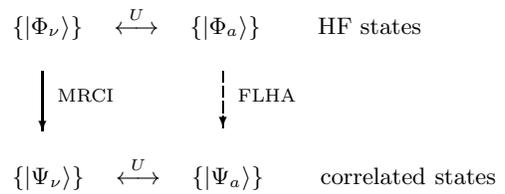

Within our frozen local hole approach we do not follow the three-step process
depicted in Fig.~\ref{figholestates} but directly generate approximate
correlated local hole states (CLHSs) $|\tilde{\Psi}_a\rangle$ by performing
separate correlation calculations for each reference state $|\Phi_a\rangle$
during which the hole in the localized orbital $|a\rangle$ is kept
frozen. This implies a configuration selection where only those configurations
are used in which the hole in $|a\rangle$ is maintained. In the case of an
CI(SD) (CI with single and double excitations) calculation the approximate
localized hole states $|\tilde{\Psi}_a\rangle$ take on the following form:
\begin{equation}
  \label{flhstates}
  |\tilde{\Psi}_a\rangle =\alpha_a |\Phi_a\rangle + \sum_{x,v} \alpha_{a,x}^v
  |\Phi_{a,x}^v\rangle + \sum_{\substack{x,x'\\v,v'}} \alpha_{a,x,x'}^{v,v'}
  |\Phi_{a,x,x'}^{v,v'}\rangle 
  \;.
\end{equation} 
The indices $x,x'\in \{\bar{a},b,\bar{b},...\}$ run over all the remaining
valence electrons, with $\bar{a}$ being the electron with opposite spin to the
removed electron $a$. $v,v'$ denote the electrons in the virtual
orbitals. Thus, e.~g., $|\Phi_{a,\bar{a}}^v\rangle$ is the 2h1p configuration
with no electrons remaining in the spatial orbital associated with $a$ whereas
$|\Phi_{a,b}^v\rangle$ contains holes in two different spatial
orbitals. Obviously, the configuration spaces for the different hole states
$|\tilde{\Psi}_a\rangle$ are overlapping. Excited configurations with holes in
two or more valence orbitals $|a\rangle $, $|b\rangle $, ... are, in fact,
present in the correlation calculations of each reference state
$|\Phi_a\rangle$, $|\Phi_b\rangle$, $\ldots$ (see Fig.~\ref{figconfig}). As we
will see in detail later, this fact does not harm the procedure. On the
contrary, it is important to build up the correlation effects in the final,
delocalized hole states properly through a suitable mixing of the
configurations in the final diagonalization step of the effective Hamiltonian
(\ref{effham}),
\begin{eqnarray}
  \label{def:Psi_tilde}
  |\tilde{\Psi}_\nu \rangle = \sum_a \lambda_a(\nu) |\tilde{\Psi}_a \rangle
  \quad\quad\mbox{where} \\ 
  \label{diagonalization}
  \sum_b H_{ab}^{\rm eff} \lambda_b(\nu) = 
  \tilde{E}_{\nu} \sum_b S_{ab} \lambda_b(\nu) \quad .
\end{eqnarray}
Note, that the individual approximated CLHSs $|\tilde{\Psi}_a\rangle$ are not
orthonormal with respect to each other. Thus, one has to solve a generalized
eigenvalue problem here (with $S_{ab} = \langle \tilde{\Psi}_a |\tilde{\Psi}_b
\rangle$ being the overlap matrix) in order to arrive at the approximate
canonical hole states $|\tilde{\Psi}_{\nu}\rangle$ and the approximate hole
state energies $\tilde{E}_{\nu}$.
\vspace{-0.3cm}

\section{Applications}
\begin{figure}
 \includegraphics[width=7.5cm,clip=true]{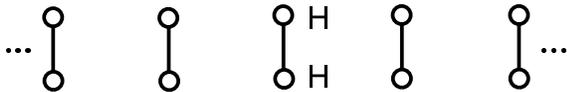}
\caption{Geometry of the (H$_2$)$_n$ ladders used in this paper. The H$_2$
bond length is fixed at its molecular value of $R=0.7417$~\AA; the distance
between the H$_2$ units is varied during the calculations.}
\label{figh2ladder}
\end{figure}

\vspace*{-0.2cm}
As first applications of the frozen local hole approximation (FLHA) we choose
two simple model systems, (H$_2$)$_n$ ladders (see Fig.~\ref{figh2ladder}) and
linear H--(Be)$_n$--H chains, for which we compare and analyze the results of
the approximation with the results of the corresponding complete
multi-reference CI(SD) (MRCI(SD)) calculation. All results are obtained by
using the {\sc MOLPRO} program package \cite{MOLPRO}, in particular the
MRCI(SD) option \cite{WK88,KW88,KW92}. For the calculations {\it sp} cc-pVDZ
basis sets by Dunning \cite{D89} are used for H and Be except for the
terminating H atoms in the Be chains which are described by a reduced {\it s}
cc-pVDZ basis. The \nobreak{Be--H} distance has been set to $1.37$~\AA\ which
is the equilibrium distance found in all neutral chains of length
$n$=4--10. The computation of the Hartree-Fock ground state of the neutral
system $|\Phi^{N}_0\rangle$ yields the canonical Hartree-Fock orbitals. The
set of valence orbitals is subsequently localized by means of the Foster-Boys
procedure \cite{FB60} and is used for the construction of the localized hole
states $\{|\Phi_{a}\rangle\}$. Each $|\Phi_{a}\rangle$ is then separately
correlated in a CI(SD) calculation with the above-described configuration
selection, where only those single and double excitations are allowed with a
hole in the localized orbital $|a\rangle$. In order to achieve this
configuration restriction within the {\sc MOLPRO} program, one has to declare
all valence and unoccupied orbitals as active orbitals. The maximum number of
active orbitals is fixed to 32 in the {\sc MOLPRO} code, which only allowed us
to handle relatively small chains and basis sets chosen here. Nevertheless,
the effects of the approximation can be studied very well in these model
systems.

\begin{figure}
\includegraphics*[width=7.5cm,clip=true]{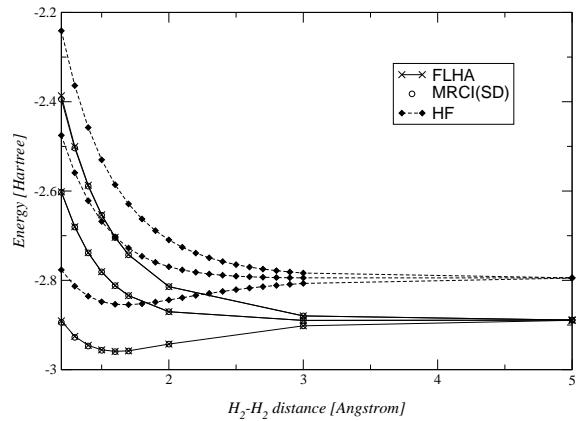}
\caption{Total Hartree-Fock, MRCI(SD) and FLHA energies of (H$_2$)$_3^+$
as a function of the H$_2$--H$_2$ distance.}
\label{figh2chainsa}
\end{figure}

\begin{figure}
\includegraphics*[width=7.5cm,clip=true]{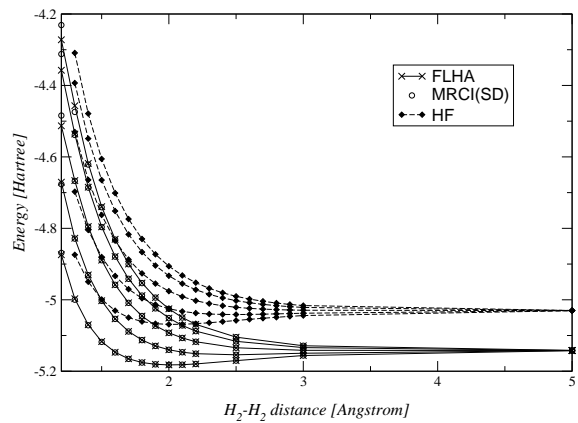}
\caption{Total Hartree-Fock, MRCI(SD) and FLHA energies of (H$_2$)$_5^+$
as a function of the H$_2$--H$_2$ distance.}
\label{figh2chainsb}
\end{figure}

The results for the (H$_2$)$_3^+$ and (H$_2$)$_5^+$ ladders are shown in
Figs.~\ref{figh2chainsa} and \ref{figh2chainsb}. Here, the total energies of
the hole states on the HF, the MRCI(SD) and the FLHA level are depicted in
dependence of the distance between the H$_2$ units. In accordance with the
number of valence orbitals, and thus possible hole states, we find three and
five low-lying cationic states of (H$_2$)$_3^+$ and (H$_2$)$_5^+$,
respectively. In their lowest states the cationic (H$_2$)$_n$ ladders are
stable. Compared to the HF data, very pronounced correlation effects are
observed: In the case of the (H$_2$)$_3^+$ chain the lowest-lying state is
lowered by about 2.9~eV at its equilibrium distance, which is slightly shifted
from $R_{\rm eq}=1.7$ to $1.6$~\AA, and the potential well is deepened from
1.6 to 1.9~eV. Also for the high-lying states of (H$_2$)$_3^+$ and the states
of (H$_2$)$_5^+$ states we find an overall energetical lowering of about 3~eV
through correlation. The most important finding in the present context,
however, is that all the described effects are very well accounted for by the
frozen local hole approximation. The FLHA curves follow the MRCI reference
data very closely with the only noticeable slight deviations at the outer left
edge of the potential curves.

\begin{figure}
\includegraphics[width=7.5cm,clip=true]{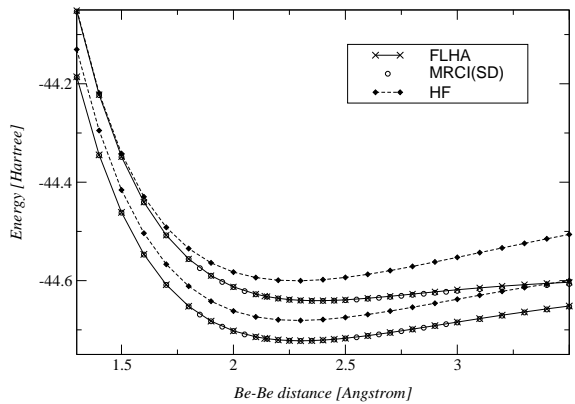}
\caption{Total Hartree-Fock, MRCI(SD) and FLHA energies for 
Be$_3$H$_2^+$ in dependence of the Be--Be distance.}
\label{figbe3h2}
\end{figure}

In order to make sure that this excellent agreement is not only an artefact of
the relatively weak van der Waals binding between the H$_2$ units, we choose
the predominantly covalently bound H--(Be)$_n$--H chains as second
application. For all systems considered, H--(Be)$_3$--H through
H--(Be)$_5$--H, the FLHA and the ``exact'' MRCI(SD) data agree very well. In
the following, we only want to discuss in detail the results for the smallest
system Be$_3$H$_2^+$ shown in Fig.~\ref{figbe3h2}. The calculated correlated
potential energy curves for the two lowest cationic states exhibit shallow
minima at Be--Be distances of $R_{\rm eq}=2.3$ and $2.4$~\AA.  For the most
stable cationic state of Be$_3$H$_2$ the correlation energies vary from about
1.1~eV at the minima to about 1.5~eV for small and large Be--Be distances.

\begin{figure}
\includegraphics[width=5.5cm,clip=true]{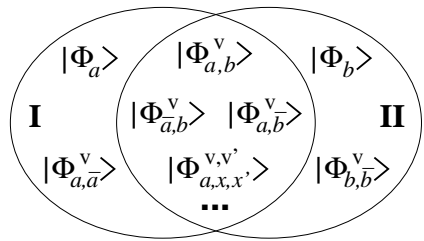}
\caption{Sketch of the overlapping configuration spaces I and II for a system
with two valence orbitals and two hole states. The nomenclature is the same as
in Eq.~(\ref{flhstates}).}
\label{figconfig}
\end{figure}

\begin{figure}
\includegraphics[width=6.5cm,clip=true]{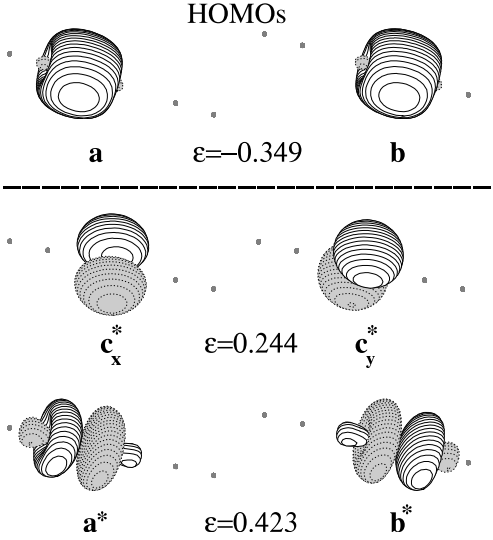}
\caption{Localized highest occupied orbitals (HOMOs) of Be$_3$H$_2^+$ together
with those localized virtual orbitals which are involved in the most relevant
configurations of the approximate correlated local hole states
$|\tilde{\Psi}_a\rangle$ and $|\tilde{\Psi}_b\rangle$. The Hartree-Fock
orbital energies $\epsilon$ of each energetically degenerate pair of orbitals
are given as well.}
\label{figvirtualorb}
\end{figure}

The dominating configurations of the two lowest cationic states are the two
reference configurations $|\Phi_a\rangle$ and $|\Phi_b\rangle$ which result
from taking one electron out of one of the highest occupied molecular orbitals
(HOMOs) of the HF determinant. The corresponding localized orbitals are
centered at the Be--Be bonds (see top row of Fig.~\ref{figvirtualorb}). Within
the FLHA two configuration spaces I and II are constructed, built upon the two
reference configurations $|\Phi_a\rangle$ and $|\Phi_b\rangle$. They are shown
schematically in Fig.~\ref{figconfig}. Clearly, the two configuration spaces
strongly overlap. For this simple example with two reference configurations
only, all mixed (single-excited) 2h1p configurations $|\Phi_{a,b}^v\rangle$,
$|\Phi_{\bar{a},b}^v\rangle$ and $|\Phi_{a,\bar{b}}^v\rangle$ and all
(doubly-excited) 3h2p configurations $|\Phi_{a,x,x'}^{v,v'}\rangle$ and
$|\Phi_{b,x,x'}^{v,v'}\rangle$ fall into the overlap region. Nevertheless,
very important configurations, especially the reference configurations
themselves, lie in only one of the spaces. That is why this very simple
example is already suited to show the main features of the
approximation. Strictly speaking, the single determinants
$|\Phi_{a,\bar{b}}^v\rangle$ only belong to space I (as the
$|\Phi_{\bar{a},b}^v\rangle$ only belong to space II). But since {\sc MOLPRO}
uses spin-adapted excitations these two types of single determinants always
come in singlet and triplet-like linear combinations
$(|\Phi_{a,\bar{b}}^v\rangle \pm |\Phi_{\bar{a},b}^v\rangle)/\sqrt{2}$ and are
thus both attributed to the overlap region I $\cap$ II.

In the present example, no substantial saving in the computational cost can be
gained from using the FLHA, since we substitute the single MRCI(SD)
calculation by two CI calculations with configuration spaces of almost the
same size as the original one. But this will change dramatically with bigger
systems because a given configuration can be found in at most three different
configuration spaces. E.~g., the double excited configuration
$|\Phi_{a,b,c}^{v,v'}\rangle$ would only be part of the configuration spaces
generated from the reference configurations $|\Phi_a\rangle$, $|\Phi_b\rangle$
and $|\Phi_c\rangle$. Once suitable cut-off criteria are introduced, the
configuration spaces of each reference configuration remain finite no matter
how large the systems becomes; and since the number of these spaces only grow
linearly with system size the FLHA directly leads to an ${\cal O}(N)$ method.

Coming back to the discussion of the Be$_3$H$_2$ results, only a few
configurations with noticeable weights ($>0.05$) are found in the approximated
correlated local hole states
$|\tilde{\Psi}_a\rangle$ and $|\tilde{\Psi}_b\rangle$
(Eq.~(\ref{flhstates})). The same
holds true for the ``full'' canonical MRCI wave functions,
$|\Psi^{\rm MRCI}_1\rangle$ and $|\Psi^{\rm MRCI}_2\rangle$. In order to enable an
analysis of the individual correlation contributions, we also localize the
virtual orbitals, although this is by no means a prerequisite of our FLHA. All
relevant configurations contain excitations into one of the four pair-wise
degenerated virtual orbitals shown in Fig.~\ref{figvirtualorb}. The first
pair, named $c_x^\ast$ and $c_y^\ast$, is localized at the central Be atom,
the other pair, $a^\ast$ and $b^\ast$, represents anti-bonding orbitals on the
two Be--Be bonds.

In Table \ref{tabconfig} we summarize the CI coefficients $\alpha_I$ and
$\beta_I$ of the most relevant contributions to the approximate CLHSs
$|\tilde{\Psi}_a\rangle$ and $|\tilde{\Psi}_b\rangle$, as well as the CI
coefficients $(\alpha_I \pm \beta_I) / \sqrt{2}$ of the approximated wave
functions $\tilde{\Psi}_\nu$ of the hole states 1 and 2 as resulting from the
diagonalization of the effective Hamiltonian, a ($2 \times 2$) matrix,
here. The latter CI coefficients are compared with those of the ``exact'' MRCI
wave functions, $\alpha_{1,I}^{\rm MRCI}$ and $\alpha_{2,I}^{\rm MRCI}$. For
all spin-adapted single determinants considered here: the reference
configurations, the single and double excitations into the anti-bonding
orbitals $a^\ast$ and $b^\ast$ and the 3h2p configurations involving the
orbitals $c_x^\ast$ and $c_y^\ast$, an excellent agreement is found between
the approximate and the ``exact'' CI coefficients of the holes states with
deviations well below 10\% in most of the cases. In contrast to our initial
expectation, no cancellation of artificial, orbital-relaxation-based
contributions in the FLHA wave functions occur. All configurations which are
important in the CLHSs also contribute significantly to (at least) one of the
final hole states $|\tilde{\Psi}_1\rangle$ and
$|\tilde{\Psi}_2\rangle$. Apparently, the ultimate diagonalization step just
mixes the individual contributions in such a way that the delocalized nature
of the hole states is recovered, precisely as one would expect for an
adiabatic approximation.
\vspace{-0.3cm}

\section{Perturbative analysis}
\vspace*{-0.2cm}
To understand, why the frozen local hole approximation (FLHA) works so well,
we switch from the MRCI level of theory to quasi-degenerate variational
perturbation theory (QDVPT) \cite{CD88} and try to find the leading
contributions in both, the ``exact'' hole state and the approximate hole
states according to the FLHA. Within QDVPT the correlated wave functions
$|\Psi_\nu\rangle$ of a system are written as
\begin{equation}
  |\Psi_\nu\rangle = \hat{\Omega} |\Psi^{\cal M}_\nu \rangle
\end{equation}
where $|\Psi^{\cal M}_\nu \rangle = \hat{P} |\Psi_\nu \rangle$ are the
projections of the correlated wave functions $|\Psi_\nu \rangle$ onto the
space ${\cal M}$ spanned by the local Hartree-Fock hole configurations
$\{|\Phi_a \rangle\}$, and $\hat{\Omega}$ is an operator which acts on the
``model space'' ${\cal M}$ and provides the full, correlated wave functions
$|\Psi_\nu \rangle$. It is assumed that the $|\Psi_\nu \rangle$ are those
eigenstates of the hole system which are dominated by the single determinants
forming the model space and, in particular, that the projections $|\Psi^{\cal
M}_\nu \rangle$ remain linear independent. In fact, \begin{equation}
\hat{\Omega} = \hat{P} + \hat{Q} \hat{\Omega} \quad\mbox{with}\quad
\hat{Q}\,\hat{\Omega} = \sum_{I,a} |\Phi_I\rangle \Omega_{Ia} \langle\Phi_a|
\end{equation}
where $\hat{Q} = \sum_I |\Phi_I \rangle \langle \Phi_I|$ is the projector onto
the orthogonal complement of the model space and the configurations $| \Phi_I
\rangle$ run over all single, double and higher excitations one can form from
the local model space configurations $\{|\Phi_a \rangle\}$.

In first order perturbation theory the wave operator $\hat{\Omega}$ is given
by \cite{CD88}
\begin{eqnarray}
  \label{def:Omega}
  \hat{\Omega} &=& \sum_c | \Phi_c \rangle \langle \Phi_c |
  \\ \nonumber &+& \sum_c \sum_{a<b,v} |\Phi_{ab}^v \rangle \, 
  \frac{ \langle v c || a b \rangle }
       { \varepsilon_v - \varepsilon_a - \varepsilon_b + \varepsilon_c } \langle
  \, \Phi_c | \quad.
\end{eqnarray}
We take the zeroth-order Hamiltonian $H^0$ to consist of the diagonal terms of
the Fock operator only, i.~e., $H^0_{ij} = \delta_{ij} \varepsilon_i$ with
$\varepsilon_i = F_{ii}$. The perturbation then contains both, the
off-diagonal terms $F_{ij} = \langle \varphi_i | \hat{F} | \varphi_j \rangle
\;,\; i\ne j$ of the Fock operator $\hat{F}$ and the usual two-electron
contributions $\langle i j || k l \rangle - \sum^{\rm occ}_a \langle i a || j
a \rangle$. Here and in the following, indices $a$, $b$, $\ldots$ are
understood to run over occupied localized orbitals while indices $v$, $w$,
$\ldots$ run over virtual orbitals. For arbitrary orbitals indices $i$, $j$,
$\ldots$ are used. The 12,12 notation
\begin{eqnarray}
  \langle i j || k l \rangle = 
  \langle i j | \hat{g} | k l \rangle - \langle i j | \hat{g} | l k \rangle
  \quad\quad\mbox{with} \\ \nonumber
  \langle i j | \hat{g} | k l \rangle = 
  \int \! \! \! \int \varphi^\ast_i(1) \varphi^\ast_j(2) \:r_{12}^{-1}\:
                     \varphi_k(1) \varphi_l(2) \: d1\,d2
\end{eqnarray}
is employed throughout. The $|\varphi_i \rangle$ are the (localized)
Hartree-Fock orbitals of the neutral $N$-electron system. Using that form of
the wave operator as {\it variational} ansatz for the eigenstates of the hole
system, i.~e.
\begin{equation}
  |\Psi_\nu \rangle
  = \hat{\Omega} \big[ \: \sum_a \lambda_a(\nu) |\Phi_a \rangle \: \big]
  = \sum_a \lambda_a(\nu) \: \hat\Omega |\Phi_a \rangle
  \quad,
\end{equation}
leads to the following effective secular matrix 
\begin{eqnarray}
  H^{\rm eff}_{ef} &:=& 
  \langle \, \hat{\Omega}\!\big[ \Phi_e \big] |\hat{H}| \,
  \hat{\Omega}\!\big[ \Phi_f \big] \rangle \\[2mm]
  &=& - F_{fe} \\ \nonumber
  && + \sum_{a<b,v} W_{e|ab,f}^{\;\;v} \: \langle v f || a b \rangle  
  \\ \nonumber
  && + \sum_{a<b,v} W_{e,ab|f}^{\;\;v} \: \langle a b || v e \rangle  
  \\ \nonumber
  && + \sum_{a<b,v} \sum_{c<d,w} W_{e,ab|cd,f}^{\;\;v\;\;\,w} \:
  \langle a b || v e \rangle \: \langle w f || c d \rangle  
  \\ \nonumber
  \mbox{with} \\
  W_{e|ab,f}^{\;\;v} &=& \frac{ \langle \Phi_e | \hat{H} | \Phi_{ab}^v \rangle }
  { ( \varepsilon_v - \varepsilon_a - \varepsilon_b + \varepsilon_f ) } 
  \\ \nonumber
  W_{e,ab|f}^{\;\;v} &=& \frac{ \langle \Phi_{ab}^v | \hat{H} | \Phi_f \rangle }
  { ( \varepsilon_v - \varepsilon_a - \varepsilon_b + \varepsilon_e ) }
  \;=\; ( W_{f|ab,e}^{\;\;\,v} )^\ast 
  \\ \nonumber
  W_{e,ab|cd,f}^{\;\;v\;\;\,w} &=& 
  \frac{ \langle \Phi_{ab}^v | \hat{H} | \Phi_{cd}^w \rangle }
  { ( \varepsilon_v - \varepsilon_a - \varepsilon_b + \varepsilon_e ) \:
  ( \varepsilon_w - \varepsilon_c - \varepsilon_d + \varepsilon_f ) }
  \;\;.
\end{eqnarray}
This is not what is usually done in QDVPT, but to preceed this way allows to
find the direct link to the frozen local hole approximation we are looking
for.

In the first step of the FLHA, the model space configurations and the excited
configurations are restricted to those determinants which already contain a
hole in a particular local occupied orbital $|\varphi_h \rangle$, such that
the wave operator reduces to
\begin{equation}
  \hat{\Omega}^{\rm FLH}(h) = |\Phi_h \rangle \langle \Phi_h |
  \: + \sum_{a \ne h,v} | \Phi_{ah}^v \rangle \,
  \frac{ \langle v h || a h \rangle }
       { \varepsilon_v - \varepsilon_a } \langle \Phi_h |
  \quad.
\label{eqomegaflh}
\end{equation}
Only single excitations $|\Phi_{ah}^v\rangle$ with respect to the given local
hole configuration $|\Phi_h\rangle$ show up here. Inspection of
Eq.~(\ref{def:Omega}) reveals that these restricted wave operators are
precisely the leading terms in the full wave operator $\hat{\Omega}$, in the
sense that
\begin{equation}
  \hat{\Omega} = \sum_h \hat{\Omega}^{\rm FLH}(h) + \hat{\Pi}
\end{equation}
where the remainder
\begin{equation}
  \hat{\Pi} = \sum_{a<b,v} \sum_{c \not\in \{a,b\}}
  |\Phi_{ab}^v\rangle \, 
  \frac{\langle v c || a b \rangle }
       {\varepsilon_v - \varepsilon_a - \varepsilon_b + \varepsilon_c }
  \, \langle\Phi_c|
\end{equation}
contains all terms in which the three occupied spin orbitals $|a\rangle$,
$|b\rangle$, and $|c\rangle$ are all {\it different}, while the leading terms
are made up by all those contributions in which two of the hole indices
coincide.

The final secular matrix of the frozen local hole approach,
\begin{equation}
  H^{\rm FLH}_{gh} := 
  \langle\,\hat{\Omega}^{\rm FLH}(g)\!\big[ \Phi_g \big] | \hat{H} |\,
  \hat{\Omega}^{\rm FLH}(h)\!\big[ \Phi_h \big] \rangle \quad,
\end{equation}
is then given by
\begin{eqnarray}
  H^{\rm FLH}_{gh} &=& - F_{hg} \\ \nonumber
  && + \sum_{c \ne h,w} W_{g|ch,h}^{\;\;w} \langle w h || c h \rangle 
  \\ \nonumber
  && + \sum_{a \ne g,v} W_{g,ag|h}^{\;\;v} \langle a g || v g \rangle 
  \\ \nonumber
  && +\sum_{a \ne g,v} \sum_{c \ne h,w} W_{g,ag|ch,h}^{\;\;v\;\;\;w}
  \langle a g || v g \rangle \langle w h || c h \rangle
\end{eqnarray}
which is again precisely the leading coinciding-hole-indices part of the full
effective Hamiltonian $H^{\rm eff}_{gh}$ (despite the always present
zeroth-order contribution $-F_{hg}$). Thus, up to first order perturbation
theory the FLHA can be understood as a simple neglection of all
three-distinct-spin-orbital contributions to the wave operator $\hat{\Omega}$
and the resulting effective Hamiltonian.

A further simplification of the FLHA can be achieved by treating the
($N$-1)-particle system with a frozen local hole on the Hartree-Fock level
only. This has already been done with some success in our ``simplified
method'' to correlation effects in band structures \cite{BFSpp}. Orbital
relaxation around the localized hole is the only effect which can be accounted
for in this approach and we want to analyze here, to which extent the
correlation effects in the ($N$-1)-particle system can be mimicked this way.

Removal of an electron from a fixed localized occupied spin orbital
$|\varphi_h\rangle$ leads to the following modified Fock operator 
$\hat{F}^{\rm FLH}(h)$,
\begin{eqnarray}\nonumber
  \langle i | \hat{F}^{\rm FLH}(h) | j \rangle
  &=& F_{ij} - \langle i h || j h \rangle
  \:=\: \delta_{ij} \, \varepsilon_i + \langle i | V | j \rangle
  \\
  \mbox{with} \quad \langle i | V | j \rangle &=& 
  (1 - \delta_{ij} ) \, F_{ij} -  \langle i h || j h \rangle
\end{eqnarray}

Up to first order in the perturbation potential $V$ the relaxed Hartree-Fock
orbitals $|\tilde{\varphi}_i(h)\rangle$ which result from diagonalizing
$\hat{F}^{\rm FLH}(h)$ under the constraint that $|\tilde{\varphi}_h\rangle$
remains unchanged read
\begin{equation}
|\tilde{\varphi}_i(h)\rangle = \left\{ \begin{array}{ll} \displaystyle{
|\varphi_i\rangle - \sum_{j \not\in\{i,h\}} |\varphi_j \rangle \,
\frac{\langle j | V | i \rangle }{\varepsilon_j - \varepsilon_i} } 
& \quad\mbox{for } i \ne h \\
|\varphi_h\rangle & \quad\mbox{else}
\end{array} \right.
\end{equation}
and the corresponding Slater determinant
\begin{eqnarray}
|\Phi_h^{\rm SCF}\rangle &:=& \frac{(-1)^{N-h}}{\sqrt{(N-1)!}} \times \\ \nonumber 
&& \mbox{det} (\tilde{\varphi}_1(h),\ldots,\tilde{\varphi}_{h-1}(h),
\tilde{\varphi}_{h+1}(h),\ldots,\tilde{\varphi}_N(h))
\end{eqnarray}
becomes
\begin{equation}
|\Phi_h^{\rm SCF}\rangle =
|\Phi_h\rangle - \sum_{a \ne h} \sum_{j \not\in\{a,h\}} |\Phi_{ah}^j\rangle \, 
\frac{\langle j | V | a \rangle }{\varepsilon_j - \varepsilon_a } + 
{\cal O}(V^2) \quad .
\end{equation}
Excitations into occupied spin orbitals $|\varphi_j\rangle $ other than
$|\varphi_a\rangle $ and $|\varphi_h\rangle $ are not possible in the 2h1p
configurations $|\Phi_{ah}^j\rangle $. Thus, $j$ only runs over the virtual
orbitals and one has
\begin{eqnarray}
|\Phi_h^{\rm SCF}\rangle &=& |\Phi_h\rangle +  
\sum_{a \ne h,v} |\Phi_{ah}^v\rangle \,
\frac{\langle v h || a h \rangle }{\varepsilon_v - \varepsilon_a} + {\cal O}(V^2)
\\ &=& \hat{\Omega}^{\rm FLH}(h) |\Phi_h\rangle + {\cal O}(V^2)
\end{eqnarray}
Obviously, up to first order, the local frozen hole wave function obtained on
the Hartree-Fock level is identical to the fully correlated local hole state
$|\tilde{\Psi}_h\rangle = \hat{\Omega}^{\rm FLH}(h) |\Phi_h\rangle$ of the
system (see Eq.~(\ref{eqomegaflh})). This result is somewhat surprising, at
first glance, because it tells, that what formally looks like pure orbital
relaxation around a frozen local defect, is actually bare electron
correlation. It also tells, that performing a Hartree-Fock calculation around
a frozen local hole rather than a much more demanding wavefunction-based
correlation calculations is a rather promising approximation.

\section{Conclusions}

The so-called frozen local hole approximation (FLHA) has been analyzed in this
study. It presents a method that allows to perform correlation calculations
for cationic and anionic ($N$$\pm$1)-electron systems in a much more efficient
way than running a full quantum chemical correlation calculation such as
MRCI(SD). In this work we were focusing on cationic holes states, but, in a
totally analogue way, the approximation can also be applied to anionic
electron attachment states (as done in Ref.~\cite{BFSpp}).

The FLHA is a two step procedure: In the first step, the electron to be
removed is assumed to reside in a given, fixed local occupied orbital and
standard wavefunction-based correlation calculations are performed in order to
find the correlation hole around each of these local holes. Because of the
frozen character of the given hole orbital the configuration space is
substantially reduced in these correlation calculations. Therefore, the
approximation is a perfect, linear scaling divide-and-conquer approach. In the
second step, the many-body Hamiltonian matrix elements with the correlated
local hole states generated in the first step are evaluated and the resulting
effective Hamiltonian matrix is diagonalized. In this way, the hole can
delocalize over the whole system to form proper Bloch states. Hence, the FLHA
can be understood as an adiabatic treatment, in which the shape of the
correlation hole around each electron hole is kept fixed during the final
hybridization of the correlated local hole states.

It has been shown numerically, for two sample systems, $({\rm H}_2)_n$ ladder
chains and hydrogen terminated linear H--(Be)$_n$--H chains, that the FLHA
performs astonishingly well. Potential energy curves have been calculated on
the MRCI(SD) level of theory, and deviations of at most 0.1 eV between a full
correlation calculations and the frozen local hole approach were found, both,
for the van der Waals bound H$_2$ chains and for the covalent Be chains.

Using quasi-degenerate variational perturbation theory, it was possible to
demonstrate that the FLHA indeed assembles all leading terms of a full
correlation calculation. Actually, the approximation consists in a neglection
of all three-distinct-hole-site contributions over those where (at least) two
hole orbitals coincide. Furthermore, it could be shown that the correlation
calculation around a frozen {\it local} hole can very well be replaced by a
numerically much less demanding SCF calculation. In first order perturbation
theory the resulting orbital relaxations are totally equivalent to the
correlation effects around a frozen local hole.

\begin{table*}
\begin{ruledtabular}
\caption{Analysis of the most relevant configurations in the two lowest
correlated cationic hole states $|\Psi_1\rangle$ and $|\Psi_2\rangle$ of
Be$_3$H$_2$. The first columns give the occupation numbers of the involved
spatial orbitals (Fig.~\ref{figvirtualorb}). The indices $S$ and $T$ refer to
spin-adapted configurations with a singlet and triplet-like linear combination
of the remaining electrons in a and b, respectively. The next two columns show
the CI coefficients $\alpha_I$ and $\beta_I$ of the approximate correlated
hole states $|\tilde{\Psi}_a\rangle$ and $|\tilde{\Psi}_b\rangle$,
respectively, as defined in Eq.~(\ref{flhstates}), followed by the CI
coefficients of the hole states $|\tilde{\Psi}_{\nu=1}\rangle$ and
$|\tilde{\Psi}_{\nu=2}\rangle$ after the diagonalization step
(\ref{diagonalization}). They are to be compared to the CI coefficients
$\alpha_{1,I}^{{\rm MRCI}}$ and $\alpha_{2,I}^{{\rm MRCI}}$ of the ``exact''
MRCI wave functions $|\Psi^{\rm MRCI}_{\nu=1}\rangle$ and $|\Psi^{\rm
MRCI}_{\nu=2}\rangle$. \\}
\begin{tabular}{llllll|cccccc|c}
\multicolumn{6}{c|}{configuration $I$} & 
$|\tilde{\Psi}_a\rangle$ & $|\tilde{\Psi}_b\rangle$ &
$|\tilde{\Psi}_1\rangle$ & $|\Psi^{\rm MRCI}_1\rangle$ & 
$|\tilde{\Psi}_2\rangle$ & $|\Psi^{\rm MRCI}_2\rangle$ & type of configuration
\\[0.8ex]
a & b & $|$ & a$^\ast$ & b$^\ast$ & c$^\ast$ & 
{\small $\alpha_I$} & 
{\small $\beta_I$} & 
{\small $\frac{1}{\sqrt{2}}(\alpha_I + \beta_I)$ } &
{\small $\alpha^{\rm MRCI}_{1,I}$} &
{\small $\frac{1}{\sqrt{2}}(\alpha_I - \beta_I) $} &
{\small $\alpha^{\rm MRCI}_{2,I}$} &
\\[0.8ex]
\hline
1&2&$|$&0&0&0     &  \phantom{-}0.9528 &  --      & \phantom{-}0.6737 &
\phantom{-}0.6767 & \phantom{-}0.6737 & \phantom{-}0.6682 & ref. configuration
$| \Phi_a \rangle$\\
2&1&$|$&0&0&0     &  --      & -0.9528 &          -0.6737  &           -0.6767
& \phantom{-}0.6737 & \phantom{-}0.6682 & ref. configuration $| \Phi_b
\rangle$
\\[0.8ex] 
1&1&$|$&1&0&0$^S$ &  \phantom{-}0.0824 & -0.0901 &           -0.0054 &           -0.0052 & \phantom{-}0.1220 & \phantom{-}0.1450 \\
1&1&$|$&1&0&0$^T$ &  \phantom{-}0.0595 & -0.1028 &           -0.1148 &           -0.1026 & \phantom{-}0.0306 & \phantom{-}0.0307 \\
1&1&$|$&0&1&0$^S$ &  \phantom{-}0.0901 & -0.0824 & \phantom{-}0.0054 &
\phantom{-}0.0052 & \phantom{-}0.1220 & \phantom{-}0.1450 & \raisebox{1.5ex}[-1.5ex]{mixed 2h1p configurations}\\
1&1&$|$&0&1&0$^T$ & -0.1028 & -0.0595 &   -0.1148 &           -0.1026 &
-0.0306 &           -0.0307 & 
\\[0.8ex]
0&2&$|$&0&1&0     &  \phantom{-}0.0507 &  --      & \phantom{-}0.0359 &
\phantom{-}0.0302 & \phantom{-}0.0359 & \phantom{-}0.0451 & type I 2h1p configurations \\
2&0&$|$&1&0&0     &  --      & -0.0507 &           -0.0359 &           -0.0302
& \phantom{-}0.0359 & \phantom{-}0.0451 & type II 2h1p configurations 
\\[0.8ex]
1&0&$|$&0&0&2     & -0.0651 & -0.0298 &           -0.0671 &           -0.0617 &           -0.0250 &          -0.0297 \\
0&1&$|$&0&0&2     &  \phantom{-}0.0298 &  \phantom{-}0.0651 & \phantom{-}0.0671 & \phantom{-}0.0617 &           -0.0250 &          -0.0297& \raisebox{1.5ex}[-1.5ex]{3h2p configurations} \\
\end{tabular}
\label{tabconfig}
\end{ruledtabular}
\end{table*}


\begin{thebibliography}{41}
\expandafter\ifx\csname natexlab\endcsname\relax\def\natexlab#1{#1}\fi
\expandafter\ifx\csname bibnamefont\endcsname\relax
  \def\bibnamefont#1{#1}\fi
\expandafter\ifx\csname bibfnamefont\endcsname\relax
  \def\bibfnamefont#1{#1}\fi
\expandafter\ifx\csname citenamefont\endcsname\relax
  \def\citenamefont#1{#1}\fi
\expandafter\ifx\csname url\endcsname\relax
  \def\url#1{\texttt{#1}}\fi
\expandafter\ifx\csname urlprefix\endcsname\relax\def\urlprefix{URL }\fi
\providecommand{\bibinfo}[2]{#2}
\providecommand{\eprint}[2][]{\url{#2}}

\bibitem[{\citenamefont{{Gr\"{a}fenstein}
  et~al.}(1993)\citenamefont{{Gr\"{a}fenstein}, Stoll, and Fulde}}]{GSF93}
\bibinfo{author}{\bibfnamefont{J.}~\bibnamefont{{Gr\"{a}fenstein}}},
  \bibinfo{author}{\bibfnamefont{H.}~\bibnamefont{Stoll}}, \bibnamefont{and}
  \bibinfo{author}{\bibfnamefont{P.}~\bibnamefont{Fulde}},
  \bibinfo{journal}{Chem.~Phys.~Lett.} \textbf{\bibinfo{volume}{215}},
  \bibinfo{pages}{611} (\bibinfo{year}{1993}).

\bibitem[{\citenamefont{{Gr\"{a}fenstein}
  et~al.}(1997)\citenamefont{{Gr\"{a}fenstein}, Stoll, and Fulde}}]{GSF97}
\bibinfo{author}{\bibfnamefont{J.}~\bibnamefont{{Gr\"{a}fenstein}}},
  \bibinfo{author}{\bibfnamefont{H.}~\bibnamefont{Stoll}}, \bibnamefont{and}
  \bibinfo{author}{\bibfnamefont{P.}~\bibnamefont{Fulde}},
  \bibinfo{journal}{Phys.~Rev.~B} \textbf{\bibinfo{volume}{55}},
  \bibinfo{pages}{13588} (\bibinfo{year}{1997}).

\bibitem[{\citenamefont{Albrecht et~al.}(2000)\citenamefont{Albrecht, Fulde,
  and Stoll}}]{AFS00}
\bibinfo{author}{\bibfnamefont{M.}~\bibnamefont{Albrecht}},
  \bibinfo{author}{\bibfnamefont{P.}~\bibnamefont{Fulde}}, \bibnamefont{and}
  \bibinfo{author}{\bibfnamefont{H.}~\bibnamefont{Stoll}},
  \bibinfo{journal}{Chem.~Phys.~Lett.} \textbf{\bibinfo{volume}{319}},
  \bibinfo{pages}{355} (\bibinfo{year}{2000}).

\bibitem[{\citenamefont{Albrecht and Igarashi}(2001)}]{AI01}
\bibinfo{author}{\bibfnamefont{M.}~\bibnamefont{Albrecht}} \bibnamefont{and}
  \bibinfo{author}{\bibfnamefont{J.-I.} \bibnamefont{Igarashi}},
  \bibinfo{journal}{J.~Phys.~Soc.~Jap.} \textbf{\bibinfo{volume}{70}},
  \bibinfo{pages}{1035} (\bibinfo{year}{2001}).

\bibitem[{\citenamefont{Albrecht}(2002)}]{A02}
\bibinfo{author}{\bibfnamefont{M.}~\bibnamefont{Albrecht}},
  \bibinfo{journal}{Theo.~Chem.~Acc.} \textbf{\bibinfo{volume}{107}},
  \bibinfo{pages}{71} (\bibinfo{year}{2002}).

\bibitem[{\citenamefont{V.Bezugly and Birkenheuer}(2004)}]{BB04}
\bibinfo{author}{\bibnamefont{V.Bezugly}} \bibnamefont{and}
  \bibinfo{author}{\bibfnamefont{U.}~\bibnamefont{Birkenheuer}},
  \bibinfo{journal}{Chem.~Phys.~Lett.} \textbf{\bibinfo{volume}{399}},
  \bibinfo{pages}{57} (\bibinfo{year}{2004}).

\bibitem[{\citenamefont{de~Graaf et~al.}(1999)\citenamefont{de~Graaf, Sousa,
  and Broer}}]{GSB99}
\bibinfo{author}{\bibfnamefont{C.}~\bibnamefont{de~Graaf}},
  \bibinfo{author}{\bibfnamefont{C.}~\bibnamefont{Sousa}}, \bibnamefont{and}
  \bibinfo{author}{\bibfnamefont{R.}~\bibnamefont{Broer}},
  \bibinfo{journal}{J.~Mol.~Struc.\ (TheoChem)} \textbf{\bibinfo{volume}{458}},
  \bibinfo{pages}{53} (\bibinfo{year}{1999}).

\bibitem[{\citenamefont{Hozoi et~al.}(2002)\citenamefont{Hozoi, de~Vries, van
  Oosten, Broer, Cabrero, and de~Graaf}}]{HVOB02}
\bibinfo{author}{\bibfnamefont{L.}~\bibnamefont{Hozoi}},
  \bibinfo{author}{\bibfnamefont{A.~H.} \bibnamefont{de~Vries}},
  \bibinfo{author}{\bibfnamefont{A.~B.} \bibnamefont{van Oosten}},
  \bibinfo{author}{\bibfnamefont{R.}~\bibnamefont{Broer}},
  \bibinfo{author}{\bibfnamefont{J.}~\bibnamefont{Cabrero}}, \bibnamefont{and}
  \bibinfo{author}{\bibfnamefont{C.}~\bibnamefont{de~Graaf}},
  \bibinfo{journal}{Phys.~Rev.~Lett.} \textbf{\bibinfo{volume}{89}},
  \bibinfo{pages}{076407} (\bibinfo{year}{2002}).

\bibitem[{\citenamefont{Hozoi et~al.}(2003)\citenamefont{Hozoi, Presura,
  de~Graaf, and Broer}}]{HPGB03}
\bibinfo{author}{\bibfnamefont{L.}~\bibnamefont{Hozoi}},
  \bibinfo{author}{\bibfnamefont{C.}~\bibnamefont{Presura}},
  \bibinfo{author}{\bibfnamefont{C.}~\bibnamefont{de~Graaf}}, \bibnamefont{and}
  \bibinfo{author}{\bibfnamefont{R.}~\bibnamefont{Broer}},
  \bibinfo{journal}{Phys.~Rev.~B} \textbf{\bibinfo{volume}{67}},
  \bibinfo{pages}{035117} (\bibinfo{year}{2003}).

\bibitem[{\citenamefont{Buth et~al.}(2005)\citenamefont{Buth, Birkenheuer,
  Albrecht, and Fulde}}]{BBAF05}
\bibinfo{author}{\bibfnamefont{C.}~\bibnamefont{Buth}},
  \bibinfo{author}{\bibfnamefont{U.}~\bibnamefont{Birkenheuer}},
  \bibinfo{author}{\bibfnamefont{M.}~\bibnamefont{Albrecht}}, \bibnamefont{and}
  \bibinfo{author}{\bibfnamefont{P.}~\bibnamefont{Fulde}},
  \bibinfo{journal}{Phys.~Rev.~B} \textbf{\bibinfo{volume}{72}},
  \bibinfo{pages}{195107} (\bibinfo{year}{2005}).

\bibitem[{\citenamefont{Pisani}(2002)}]{P02}
\bibinfo{author}{\bibfnamefont{C.}~\bibnamefont{Pisani}},
  \bibinfo{journal}{J.~Mol.~Struct.~(TheoChem)} \textbf{\bibinfo{volume}{621}},
  \bibinfo{pages}{141} (\bibinfo{year}{2002}).

\bibitem[{\citenamefont{Saeb{\o} and Pulay}(1987)}]{SP87}
\bibinfo{author}{\bibfnamefont{S.}~\bibnamefont{Saeb{\o}}} \bibnamefont{and}
  \bibinfo{author}{\bibfnamefont{P.}~\bibnamefont{Pulay}},
  \bibinfo{journal}{J.~Chem.~Phys.} \textbf{\bibinfo{volume}{86}},
  \bibinfo{pages}{914} (\bibinfo{year}{1987}).

\bibitem[{\citenamefont{Hampel and Werner}(1996)}]{HW96}
\bibinfo{author}{\bibfnamefont{C.}~\bibnamefont{Hampel}} \bibnamefont{and}
  \bibinfo{author}{\bibfnamefont{H.-J.} \bibnamefont{Werner}},
  \bibinfo{journal}{J.~Chem.~Phys.} \textbf{\bibinfo{volume}{104}},
  \bibinfo{pages}{6286} (\bibinfo{year}{1996}).

\bibitem[{\citenamefont{{Sch\"{u}tz} et~al.}(1999)\citenamefont{{Sch\"{u}tz},
  Hetzer, and Werner}}]{SHW99}
\bibinfo{author}{\bibfnamefont{M.}~\bibnamefont{{Sch\"{u}tz}}},
  \bibinfo{author}{\bibfnamefont{G.}~\bibnamefont{Hetzer}}, \bibnamefont{and}
  \bibinfo{author}{\bibfnamefont{H.-J.} \bibnamefont{Werner}},
  \bibinfo{journal}{J.~Chem.~Phys.} \textbf{\bibinfo{volume}{111}},
  \bibinfo{pages}{5691} (\bibinfo{year}{1999}).

\bibitem[{\citenamefont{Sun and Bartlett}(1996)}]{SB96}
\bibinfo{author}{\bibfnamefont{J.-Q.} \bibnamefont{Sun}} \bibnamefont{and}
  \bibinfo{author}{\bibfnamefont{R.~J.} \bibnamefont{Bartlett}},
  \bibinfo{journal}{J.~Chem.~Phys.} \textbf{\bibinfo{volume}{104}},
  \bibinfo{pages}{8553} (\bibinfo{year}{1996}).

\bibitem[{\citenamefont{Ayala et~al.}(2001)\citenamefont{Ayala, Kudin, and
  Scuseria}}]{AKS01}
\bibinfo{author}{\bibfnamefont{P.~Y.} \bibnamefont{Ayala}},
  \bibinfo{author}{\bibfnamefont{K.~N.} \bibnamefont{Kudin}}, \bibnamefont{and}
  \bibinfo{author}{\bibfnamefont{G.~E.} \bibnamefont{Scuseria}},
  \bibinfo{journal}{J.~Chem.~Phys.} \textbf{\bibinfo{volume}{115}},
  \bibinfo{pages}{9698} (\bibinfo{year}{2001}).

\bibitem[{\citenamefont{Pisani et~al.}(2005)\citenamefont{Pisani, Busse,
  Capecchi, Casassa, Dovesi, Maschio, Zicovich-Wilson, and
  {Sch\"{u}tz}}}]{PBCC05}
\bibinfo{author}{\bibfnamefont{C.}~\bibnamefont{Pisani}},
  \bibinfo{author}{\bibfnamefont{M.}~\bibnamefont{Busse}},
  \bibinfo{author}{\bibfnamefont{G.}~\bibnamefont{Capecchi}},
  \bibinfo{author}{\bibfnamefont{S.}~\bibnamefont{Casassa}},
  \bibinfo{author}{\bibfnamefont{R.}~\bibnamefont{Dovesi}},
  \bibinfo{author}{\bibfnamefont{L.}~\bibnamefont{Maschio}},
  \bibinfo{author}{\bibfnamefont{C.}~\bibnamefont{Zicovich-Wilson}},
  \bibnamefont{and}
  \bibinfo{author}{\bibfnamefont{M.}~\bibnamefont{{Sch\"{u}tz}}},
  \bibinfo{journal}{J.~Chem.~Phys.} \textbf{\bibinfo{volume}{122}},
  \bibinfo{pages}{094113} (\bibinfo{year}{2005}).

\bibitem[{\citenamefont{Fink and Staemmler}(1995)}]{FS95}
\bibinfo{author}{\bibfnamefont{K.}~\bibnamefont{Fink}} \bibnamefont{and}
  \bibinfo{author}{\bibfnamefont{V.}~\bibnamefont{Staemmler}},
  \bibinfo{journal}{J.~Chem.~Phys.} \textbf{\bibinfo{volume}{103}},
  \bibinfo{pages}{2603} (\bibinfo{year}{1995}).

\bibitem[{\citenamefont{Kitaura et~al.}(1999)\citenamefont{Kitaura, Ikeo,
  Asada, Nakano, and Uebayasi}}]{KIANU99}
\bibinfo{author}{\bibfnamefont{K.}~\bibnamefont{Kitaura}},
  \bibinfo{author}{\bibfnamefont{E.}~\bibnamefont{Ikeo}},
  \bibinfo{author}{\bibfnamefont{T.}~\bibnamefont{Asada}},
  \bibinfo{author}{\bibfnamefont{T.}~\bibnamefont{Nakano}}, \bibnamefont{and}
  \bibinfo{author}{\bibfnamefont{M.}~\bibnamefont{Uebayasi}},
  \bibinfo{journal}{Chem.~Phys.~Lett.} \textbf{\bibinfo{volume}{313}},
  \bibinfo{pages}{701} (\bibinfo{year}{1999}).

\bibitem[{\citenamefont{Fedorov and Kitaura}(2004)}]{FK04}
\bibinfo{author}{\bibfnamefont{D.~G.} \bibnamefont{Fedorov}} \bibnamefont{and}
  \bibinfo{author}{\bibfnamefont{K.}~\bibnamefont{Kitaura}},
  \bibinfo{journal}{J.~Chem.~Phys.} \textbf{\bibinfo{volume}{121}},
  \bibinfo{pages}{2483} (\bibinfo{year}{2004}).

\bibitem[{\citenamefont{Mochizuki et~al.}(2005)\citenamefont{Mochizuki,
  Koikegami, Amari, Segawa, Kitaura, and Nakano}}]{MKAS05}
\bibinfo{author}{\bibfnamefont{Y.}~\bibnamefont{Mochizuki}},
  \bibinfo{author}{\bibfnamefont{S.}~\bibnamefont{Koikegami}},
  \bibinfo{author}{\bibfnamefont{S.}~\bibnamefont{Amari}},
  \bibinfo{author}{\bibfnamefont{K.}~\bibnamefont{Segawa}},
  \bibinfo{author}{\bibfnamefont{K.}~\bibnamefont{Kitaura}}, \bibnamefont{and}
  \bibinfo{author}{\bibfnamefont{T.}~\bibnamefont{Nakano}},
  \bibinfo{journal}{Chem.~Phys.~Lett.)} \textbf{\bibinfo{volume}{406}},
  \bibinfo{pages}{283} (\bibinfo{year}{2005}).

\bibitem[{\citenamefont{Stoll}(1992{\natexlab{a}})}]{S92-1}
\bibinfo{author}{\bibfnamefont{H.}~\bibnamefont{Stoll}},
  \bibinfo{journal}{Chem.~Phys.~Lett.} \textbf{\bibinfo{volume}{191}},
  \bibinfo{pages}{548} (\bibinfo{year}{1992}{\natexlab{a}}).

\bibitem[{\citenamefont{Stoll}(1992{\natexlab{b}})}]{S92-2}
\bibinfo{author}{\bibfnamefont{H.}~\bibnamefont{Stoll}},
  \bibinfo{journal}{Phys.~Rev.~B)} \textbf{\bibinfo{volume}{B46}},
  \bibinfo{pages}{6700} (\bibinfo{year}{1992}{\natexlab{b}}).

\bibitem[{\citenamefont{Stoll}(1992{\natexlab{c}})}]{S92-3}
\bibinfo{author}{\bibfnamefont{H.}~\bibnamefont{Stoll}},
  \bibinfo{journal}{J.~Chem.~Phys.} \textbf{\bibinfo{volume}{97}},
  \bibinfo{pages}{8449} (\bibinfo{year}{1992}{\natexlab{c}}).

\bibitem[{\citenamefont{Paulus et~al.}(1995)\citenamefont{Paulus, Fulde, and
  Stoll}}]{PFS95}
\bibinfo{author}{\bibfnamefont{B.}~\bibnamefont{Paulus}},
  \bibinfo{author}{\bibfnamefont{P.}~\bibnamefont{Fulde}}, \bibnamefont{and}
  \bibinfo{author}{\bibfnamefont{H.}~\bibnamefont{Stoll}},
  \bibinfo{journal}{Phys.~Rev.~B} \textbf{\bibinfo{volume}{51}},
  \bibinfo{pages}{10572} (\bibinfo{year}{1995}).

\bibitem[{\citenamefont{Fulde}(2002)}]{F02}
\bibinfo{author}{\bibfnamefont{P.}~\bibnamefont{Fulde}},
  \bibinfo{journal}{Adv.~Phys.} \textbf{\bibinfo{volume}{51}},
  \bibinfo{pages}{909} (\bibinfo{year}{2002}).

\bibitem[{\citenamefont{Willnauer and Birkenheuer}(2004)}]{WB04}
\bibinfo{author}{\bibfnamefont{C.}~\bibnamefont{Willnauer}} \bibnamefont{and}
  \bibinfo{author}{\bibfnamefont{U.}~\bibnamefont{Birkenheuer}},
  \bibinfo{journal}{J.~Chem.~Phys.} \textbf{\bibinfo{volume}{120}},
  \bibinfo{pages}{11910} (\bibinfo{year}{2004}).

\bibitem[{\citenamefont{Angeli et~al.}(2003)\citenamefont{Angeli, Calzado,
  Cimiraglia, Evangelisti, Guih{\'e}ry, Leininger, Malrieu, Maynau, Ruiz, and
  Sparta}}]{ACCE03}
\bibinfo{author}{\bibfnamefont{C.}~\bibnamefont{Angeli}},
  \bibinfo{author}{\bibfnamefont{C.~J.} \bibnamefont{Calzado}},
  \bibinfo{author}{\bibfnamefont{R.}~\bibnamefont{Cimiraglia}},
  \bibinfo{author}{\bibfnamefont{S.}~\bibnamefont{Evangelisti}},
  \bibinfo{author}{\bibfnamefont{N.}~\bibnamefont{Guih{\'e}ry}},
  \bibinfo{author}{\bibfnamefont{T.}~\bibnamefont{Leininger}},
  \bibinfo{author}{\bibfnamefont{J.-P.} \bibnamefont{Malrieu}},
  \bibinfo{author}{\bibfnamefont{D.}~\bibnamefont{Maynau}},
  \bibinfo{author}{\bibfnamefont{J.~V.~P.} \bibnamefont{Ruiz}},
  \bibnamefont{and} \bibinfo{author}{\bibfnamefont{M.}~\bibnamefont{Sparta}},
  \bibinfo{journal}{Mol.~Phys.} \textbf{\bibinfo{volume}{101}},
  \bibinfo{pages}{1389} (\bibinfo{year}{2003}).

\bibitem[{\citenamefont{Borini et~al.}(2005)\citenamefont{Borini, Maynau, and
  Evangelisti}}]{BME05}
\bibinfo{author}{\bibfnamefont{S.}~\bibnamefont{Borini}},
  \bibinfo{author}{\bibfnamefont{D.}~\bibnamefont{Maynau}}, \bibnamefont{and}
  \bibinfo{author}{\bibfnamefont{S.}~\bibnamefont{Evangelisti}},
  \bibinfo{journal}{J.~Comput.~Chem.} \textbf{\bibinfo{volume}{26}},
  \bibinfo{pages}{1042} (\bibinfo{year}{2005}).

\bibitem[{\citenamefont{Birkenheuer et~al.}(2005)\citenamefont{Birkenheuer,
  Fulde, and Stoll}}]{BFSpp}
\bibinfo{author}{\bibfnamefont{U.}~\bibnamefont{Birkenheuer}},
  \bibinfo{author}{\bibfnamefont{P.}~\bibnamefont{Fulde}}, \bibnamefont{and}
  \bibinfo{author}{\bibfnamefont{H.}~\bibnamefont{Stoll}},
  \bibinfo{journal}{Theo.~Chem.~Acc.}  (\bibinfo{year}{2005}),
  \bibinfo{note}{submitted, {\tt arXiv:cond-mat/0511626}}.

\bibitem[{\citenamefont{{R\"{o}ssler} and Staemmler}(2003)}]{RS03}
\bibinfo{author}{\bibfnamefont{N.}~\bibnamefont{{R\"{o}ssler}}}
  \bibnamefont{and}
  \bibinfo{author}{\bibfnamefont{V.}~\bibnamefont{Staemmler}},
  \bibinfo{journal}{Phys.~Chem.~Chem.~Phys.} \textbf{\bibinfo{volume}{5}},
  \bibinfo{pages}{3580} (\bibinfo{year}{2003}).

\bibitem[{\citenamefont{de~Graaf et~al.}(1997)\citenamefont{de~Graaf, Broer,
  Nieuwpoort, and Bagus}}]{GBNB97}
\bibinfo{author}{\bibfnamefont{C.}~\bibnamefont{de~Graaf}},
  \bibinfo{author}{\bibfnamefont{R.}~\bibnamefont{Broer}},
  \bibinfo{author}{\bibfnamefont{W.~C.} \bibnamefont{Nieuwpoort}},
  \bibnamefont{and} \bibinfo{author}{\bibfnamefont{P.~S.} \bibnamefont{Bagus}},
  \bibinfo{journal}{Chem.~Phys.~Lett.} \textbf{\bibinfo{volume}{272}},
  \bibinfo{pages}{341} (\bibinfo{year}{1997}).

\bibitem[{\citenamefont{de~Vries et~al.}(2002)\citenamefont{de~Vries, Hozoi,
  Broer, and Bagus}}]{VHBB02}
\bibinfo{author}{\bibfnamefont{A.~H.} \bibnamefont{de~Vries}},
  \bibinfo{author}{\bibfnamefont{L.}~\bibnamefont{Hozoi}},
  \bibinfo{author}{\bibfnamefont{R.}~\bibnamefont{Broer}}, \bibnamefont{and}
  \bibinfo{author}{\bibfnamefont{P.~S.} \bibnamefont{Bagus}},
  \bibinfo{journal}{Phys.~Rev.~B} \textbf{\bibinfo{volume}{66}},
  \bibinfo{pages}{035108} (\bibinfo{year}{2002}).

\bibitem[{\citenamefont{Koopmans}(1933)}]{K33}
\bibinfo{author}{\bibfnamefont{T.}~\bibnamefont{Koopmans}},
  \bibinfo{journal}{Physica} \textbf{\bibinfo{volume}{1}}, \bibinfo{pages}{104}
  (\bibinfo{year}{1933}).

\bibitem[{\citenamefont{Werner et~al.}(2003)\citenamefont{Werner, Knowles,
  Lindh, {Sch\"{u}tz}, Celani, Korona, Manby, Rauhut, Amos, Bernhardsson
  et~al.}}]{MOLPRO}
\bibinfo{author}{\bibfnamefont{H.-J.} \bibnamefont{Werner}},
  \bibinfo{author}{\bibfnamefont{P.~J.} \bibnamefont{Knowles}},
  \bibinfo{author}{\bibfnamefont{R.}~\bibnamefont{Lindh}},
  \bibinfo{author}{\bibfnamefont{M.}~\bibnamefont{{Sch\"{u}tz}}},
  \bibinfo{author}{\bibfnamefont{P.}~\bibnamefont{Celani}},
  \bibinfo{author}{\bibfnamefont{T.}~\bibnamefont{Korona}},
  \bibinfo{author}{\bibfnamefont{F.~R.} \bibnamefont{Manby}},
  \bibinfo{author}{\bibfnamefont{G.}~\bibnamefont{Rauhut}},
  \bibinfo{author}{\bibfnamefont{R.~D.} \bibnamefont{Amos}},
  \bibinfo{author}{\bibfnamefont{A.}~\bibnamefont{Bernhardsson}},
  \bibnamefont{et~al.}, \emph{\bibinfo{title}{Molpro, version 2002.6, a package
  of ab initio programs}} (\bibinfo{year}{2003}), \bibinfo{note}{see
  http://www.molpro.net}.

\bibitem[{\citenamefont{Werner and Knowles}(1988)}]{WK88}
\bibinfo{author}{\bibfnamefont{H.-J.} \bibnamefont{Werner}} \bibnamefont{and}
  \bibinfo{author}{\bibfnamefont{P.~J.} \bibnamefont{Knowles}},
  \bibinfo{journal}{J.~Chem.~Phys.} \textbf{\bibinfo{volume}{89}},
  \bibinfo{pages}{5803} (\bibinfo{year}{1988}).

\bibitem[{\citenamefont{Knowles and Werner}(1988)}]{KW88}
\bibinfo{author}{\bibfnamefont{P.~J.} \bibnamefont{Knowles}} \bibnamefont{and}
  \bibinfo{author}{\bibfnamefont{H.-J.} \bibnamefont{Werner}},
  \bibinfo{journal}{Chem.~Phys.~Letters} \textbf{\bibinfo{volume}{145}},
  \bibinfo{pages}{514} (\bibinfo{year}{1988}).

\bibitem[{\citenamefont{Knowles and Werner}(1992)}]{KW92}
\bibinfo{author}{\bibfnamefont{P.~J.} \bibnamefont{Knowles}} \bibnamefont{and}
  \bibinfo{author}{\bibfnamefont{H.-J.} \bibnamefont{Werner}},
  \bibinfo{journal}{Theor.~Chim.~Acta} \textbf{\bibinfo{volume}{84}},
  \bibinfo{pages}{95} (\bibinfo{year}{1992}).

\bibitem[{\citenamefont{T.~H.~Dunning}(1989)}]{D89}
\bibinfo{author}{\bibfnamefont{J.}~\bibnamefont{T.~H.~Dunning}},
  \bibinfo{journal}{J.~Chem.~Phys.} \textbf{\bibinfo{volume}{90}},
  \bibinfo{pages}{1007} (\bibinfo{year}{1989}).

\bibitem[{\citenamefont{Foster and Boys}(1960)}]{FB60}
\bibinfo{author}{\bibfnamefont{J.~M.} \bibnamefont{Foster}} \bibnamefont{and}
  \bibinfo{author}{\bibfnamefont{S.~F.} \bibnamefont{Boys}},
  \bibinfo{journal}{Rev.~Mod.~Phys.} \textbf{\bibinfo{volume}{32}},
  \bibinfo{pages}{300} (\bibinfo{year}{1960}).

\bibitem[{\citenamefont{Cave and Davidson}(1988)}]{CD88}
\bibinfo{author}{\bibfnamefont{R.~J.} \bibnamefont{Cave}} \bibnamefont{and}
  \bibinfo{author}{\bibfnamefont{E.~R.} \bibnamefont{Davidson}},
  \bibinfo{journal}{J.~Chem.~Phys.} \textbf{\bibinfo{volume}{89}},
  \bibinfo{pages}{6798} (\bibinfo{year}{1988}).

\end{thebibliography}
\end{document}